\documentclass[floatfix,aps,nofootinbib]{revtex4}
\usepackage{epsfig}

\def\lp {\left( }
\def\rp {\right) }
\def\lb {\left[ }
\def\rb {\right] }

\def\nn {\nonumber}

\def\beq{\begin{equation}}
\def\eeq{\end{equation}}
\def\bea{\begin{eqnarray}}
\def\eea{\end{eqnarray}}
\def\ni{\noindent}

\def\m{\mu}
\def\n{\nu}

\begin{document}

\title{Elements stability: a microcosmos effect?}

\author{C.C. Barros Jr.}

\affiliation{Instituto de F\'{\i}sica, Universidade de S\~{a}o Paulo,\\
C.P. 66318, 05315-970, S\~ao Paulo, SP, Brazil}


\begin{abstract}
The electronic structure of heavy elements, when described in a space-time 
which the metric is affected by the electromagnetic interaction,
presents instabilities. These instabilities increase with the atomic number, 
and above a critical value,  become important.
The consistency of this theory \cite{cb1},\cite{cb2} and a formulation based
 on the
energy-momentum tensor  is also  investigated. With this 
procedure, a dynamical cut-off appears in a natural way,
and the field equations for the general quantum mechanics
are determined.
\end{abstract}

\maketitle

\vspace{5mm}

\section{introduction}

A fundamental question in the understanding of the Nature is if large-scale
systems, such as stars or galaxies are submitted to the same principles
that rules the microscopic world. Should these small systems be considered
as microcosmos?

The microscopic world, is known to be explained in terms of the
quantum mechanics and a key element for its development was
the study of the atomic structure. The success of the quantum theory in 
studying these systems,  confirmed  
the foundations of the theory
and allowed the description of a
large number of systems such as molecules and nuclei.
The understanding of the physical reality by the quantum mechanics 
has been greatly improved with the  
 Dirac theory
\cite{dir1}, that formulated the quantum mechanics in the 
framework
 of the special relativity with remarkable success.

On the other hand, the Einstein general theory of relativity is another great
achievement, that deals with the structure of the space-time, 
and when applied to very large systems, 
 gives very precise results also.
Since 1930, many works (as for example \cite{rfd}-\cite{hawk1}) have been 
proposed 
with the objective of quantizing the gravity. Looking at these questions from
a different point of view, 
a question that must remain in many heads is if the general
relativity formulation may have
 effects, or at least, some analogy in the study of 
 very small systems, such as atoms or elementary
particles. 

In practical terms, one aspect of 
this question is if the microscopic world
interactions, the weak, electromagnetic and the strong, may affect 
significantly the space-time metric and if this proposition may have any
observable effect. In this description, the gravitational forces may be 
neglected, and it is a good approximation,
due to the small masses of the considered particles.
 In \cite{cb1}, these ideas 
have been formulated  
considering a particle in a region with a potential, that affects the metric,
 and the wave equations for spin-0 and
spin-1/2 particles
 have been proposed,  generating very interesting results. 

The simplest systems where this theory could be tested  are the one 
electron atoms, and the calculation of the deuterium spectrum has shown
a clear numerical improvement when compared with the usual
Dirac spectrum \cite{dir1}, \cite{scf} (also proposed by Sommerfeld 
\cite{som}),  with a percentual
 deviation from the experimental results 
approximately five times 
smaller, near
  one additional digit of precision.

An interesting fact that appeared from this theory, is the existence of 
horizons of events
 inside these quantum systems, 
 with sizes that are not negligible. This propriety, that is related with 
the existence of a
trapping surface at $r_0$, as it was defined by Penrose \cite{pen},
 in \cite{cb2}
have been successfully used in order to describe quark confinement.
Solving the quantum wave equations \cite{cb2},  quark confinement
has been obtained, without the need of introducing confining potentials, as it 
is currently done \cite{bog}- \cite{zz}. The confinement obtained in this way 
is a strong confinement,  
as the quarks cannot reach $r_0$.

In order to compliment this theory,
 the energy-momentum tensor $T^{\m\n}$
 should be included
in this formulation. This study is made in Sect. 2 and consistence with the
previous results is found.  
In Sect. 3, the effect of instability in heavy elements is studied, and in 
Sect. 4, the conclusions are shown.

\section{Field equations and metric}

In quantum 
systems,  the electromagnetic and strong interactions dominate and
the gravitational interaction is negligible, as the masses of the 
considered particles are very small. Consequently, 
 the gravitational
potential may be turned off, and then, 
  the space-time metric is  affected only by the other 
interactions. 

In this section, the field equations for particles subjected to 
non gravitational interactions will be obtained. For this purpose  
  a brief review of the results of \cite{cb1} will be made, 
 and  then, this results will be related  with
a formulation based on the energy-momentum tensor.

For simplicity,  a system with spherical symmetry will be considered,
 but the 
basic ideas can be generalized to systems with arbitrary metrics. 
If the spherical symmetry is considered, 
 the space-time may be described 
by the  metric derived in \cite{cb1}, that is very similar to the
Schwarzschild metric \cite{lan},\cite{wein},
\beq
ds^2=\xi\ d\tau^2 - r^2(d\theta^2+ \sin^2 \theta\ d\phi^2) - \xi^{-1}dr^2  \ 
 \  ,
\label{I.1}
\eeq 

\ni
where $r$,  $\theta$ and $\phi$ are the particle coordinates,
$\xi(r)$ is 
determined by the interaction potential $V(r)$, and is a 
function only of 
$r$, for a time independent interaction. 
In this case, the metric tensor $g_{\m\n}$ is diagonal 
\beq
g_{\m\n}=\pmatrix{\xi&0&0&0\cr 0&-\xi^{-1}&0&0\cr  0&0&-r^2&0\cr
0&0&0&-r^2\sin^2\theta\cr}  \  \  .
\label{gmn}
\eeq

 The energy relation for this system is \cite{cb1} 
\beq
{E\over \sqrt{\xi}}=\sqrt{p^2 c^2 + m_0^2 c^4}  \  \ ,
\label{en}
\eeq

\ni
therefore,
\beq
E(\vec\beta=0)=E_0\xi^{1/2}=E_0+V  \  \  ,
\eeq

\ni
where
$\vec\beta$ is the particle velocity. This relation means  that
in the rest frame of the particle, the energy is simply due to 
the sum of its 
rest mass $E_0$ with  the potential, and then, 
\beq
\xi^{1\over 2} = 1+{V\over mc^2}  \  \  .
\label{ximet}
\eeq

\ni
Applying these ideas in the study of one electron atoms, $V$ is
the Coulomb potential
\beq
V(r)=-{\alpha \ Z\over r}  \  ,
\eeq

\ni
$\alpha$ is the fine structure constant and $Z$ is the atomic 
number. Consequently, the function
 $\xi$ is given by
\beq
\xi =1 -{2\alpha \ Z\over mc^2r}+{\alpha^2Z^2\over m^2c^4r^2} \  ,
\label{xi1}
\eeq

\ni
where $m$ is the electron mass.
These expressions determine
 the horizon of events at $r_0$, that appears from the
metric singularity $\xi(r_0)$=0, and using the values of \cite{pdg},
one finds
\beq
r_0= {\alpha Z\over mc^2}=2.818\ Z \  {\rm fm}  \  \  ,
\label{r0}
\eeq

\ni that is not a negligible value at the atomic scale.

Now, let us turn our attention to a description based on the energy-momentum
tensor. 
If one consider a field generated by
  the electromagnetic interaction, the energy-momentum tensor is
\beq
T^{\m\n}= \epsilon_0 \lp F^{\m\alpha} F_{\alpha}^\n -{1\over 4} 
\delta^{\m\n} F_{\alpha\beta}F^{\alpha\beta}  \rp  \   ,
\eeq

\ni
and it is related to the space-time geometry by 
the field equations
\beq
R_\n^\m-{1\over 2}R\delta_\n^\m=-AT_\n^\m  \  ,
\label{curv}
\eeq

\ni
where $A$ is a constant to be determined.

\ni
In an  atom, 
$T^{\m\n}$ is determined by the nuclear electrostatic field, 
and the nonvanishing components are
\beq
T_0^0= {1\over 2} \lp \epsilon_0E^2+{B^2\over \m_0}   \rp  \   , 
\eeq

\ni
and $T_r^r=-T_0^0$.
For a central electrostatic field with charge $Ze$, $T_0^0$ 
is just
\beq
T_0^0={\epsilon_0E^2\over 2}= {Z^2\alpha\over 8\pi r^4}  \  \  .
\label{t00}
\eeq

Eq. (\ref{curv}) may be used to determine the $\xi$ function
observing that in the given metric
\bea
&&{e^{-B}\over r^2} \lp r{dB\over dr}-1   \rp +{1\over r^2}=AT_0^0  \nn  \\
&&{e^{-B}\over r^2} \lp r{dB\over dr}+1   \rp -{1\over r^2}=-AT_0^0  \  \  ,
\label{ein1}
\eea

\ni
that have the solution \cite{mmal}
\beq
\xi(r)=e^{-B}=1-{c^2AM(r)\over 4\pi r}
\eeq

\ni
with
\beq
M(r)=\int_0^r{4\pi\over c^2} (r')^2 T_0^0 dr'   \   .
\label{mrr}
\eeq

\ni
If the particle is outside the horizon of events, that is the case of the 
electron
in an atom, it will be affected just by the part of the field located
in the region  external  to the horizon of events, and the integration 
(\ref{mrr}) must be performed in the interval $r_0\leq r'\leq r$,
\beq
M(r)=m_0+\int_{r_0}^r {4\pi\over c^2} \ (r')^2 T_0^0(r') dr'=
m_0+{Z^2\alpha\over 2\ c^2}
\lp{1\over r_0}-{1\over r}\rp  \  ,
\label{mmm}
\eeq
\ni
where $m_0$ is a constant of integration. So,
\bea
&& M(r) = m_0 + {mZ\over 2} -{Z^2\alpha\over 2\ c^2r}    \\
&& M(r\rightarrow\infty)=m_0+ {mZ\over 2} 
\label{mm}
\eea

\ni
and
\beq
\xi= 1-{c^2AZ\over 4\pi r}\lb m_0+ {mZ\over 2} \rb + 
{AZ^2\alpha\over 8\pi r^2}
\label{mr}
\eeq 

\ni
that is solution of (\ref{ein1}).
The constants $A$ and $m_0$ may be obtained now,  comparing the expression 
(\ref{mr}), that is determined by the energy-momentum tensor and the
field equations, with (\ref{xi1}), determined by the energy relation 
(\ref{en}).

Identifying the terms $r^{-1}$ and $r^{-2}$ we find
\bea
A&=&{8\pi\alpha\over m^2c^4}   \nn \\
m_0&=&{mZ\over 2}   \   .
\eea

\ni
The field equation is then
\beq
R_\n^\m-{1\over 2}R\delta_\n^\m=-{8\pi\alpha\over m^2c^2}T_\n^\m  \  .
\label{bfld}
\eeq

So, in this system, due to the interaction,
the electron energy decrease with $r$, until it reaches the 
value $E(r_0)=0$ at the horizon of events. At large distances, $E(\infty)=m$,
the energy is obviously due only to its rest mass.

Considering that the strong interactions may be approximated by a
strong Coulomb field \cite{cb2}, with $\alpha\sim 1$, one may use 
the field equation
 (\ref{bfld}) in order to study strongly interacting systems.
 But if one wants to
 describe the strong interactions
by the Yang-Mills field, (\ref{bfld}) is determined by the correct
coupling constant $\alpha$, and an Yang-Mills energy-momentum tensor.


\section{Z Limit}

The quantum wave equation
 for spin-1/2 particles, in the  metric (\ref{I.1}) is \cite{cb1}
\beq
{i\hbar \over \xi}{\partial\over\partial t}\Psi=\lp -i\hbar c\ 
\vec \alpha.\vec\nabla +\beta m_0c^2  \rp\Psi \  \  
\label{bar}
\eeq

\ni
where $\Psi$ is a four-component spinor. In a first view, this equation may 
seem very similar to the Dirac equation, but it shows some important 
differences. The first one, is the numerical aspect, 
applying this theory to the deuterium atom \cite{cb1}, the accord with the 
experimental values of the energy levels
 is improved in comparison with the Dirac spectrum  
 and shows deviations from the data
 of the
order of 0.005\%.
Another characteristic of the general quantum mechanics, is the existence of 
the horizon of events at 
$r_0$. In the Hydrogen atom, this fact is not important, $r_0\sim$2.8 fm,
and, from the solution of (\ref{bar}),
 an electron with energy of the order of few eV has
 a very small probability of being found in this region,
so, in practical terms, no effect is 
observed.
However, for heavy elements, that present larger nuclear charges, 
the value of $r_0$ may not be  negligible as it increases with $Z$. 
For example, for the mercury, $Z=$80, so, $r_0$=224 fm, should this value be 
important?
In this section, the main objective is to study this effect.

An estimate of 
the
spacial region where the 1$S_{1/2}$ electron has significant probability of
 being
found is the interval $r_{min}=\langle r\rangle -\Delta r \leq r \leq 
\langle r\rangle+\Delta r$,
where $\langle r \rangle$ is the expectation value of $r$
and  $\Delta r=\sqrt{\langle r^2\rangle -\langle r\rangle}$
is the statistical fluctuation of $r$ around $\langle r\rangle$.
 From the solution 
of eq. (\ref{bar}) \cite{cb1}, \cite{cb2}, one may
determine theses quantities.

Solving eq. (\ref{bar}), the spacial part of $\Psi$, that is a four component
spinor, may be written as
\beq
\psi=\pmatrix{F(r)\chi_\kappa^\m\cr iG(r)\chi_{-\kappa}^\m}  \  \  ,
\eeq

\ni
with the aid of the usual two component spinors, $\chi_\kappa^\m$ \cite{cb1}, 
where
\bea
&& \kappa=l \ \  \  \  \   \  \   \  \  {\rm for} \  \  j=l-1/2 \  \  , \nn \\
&& \kappa=-l-1  \  \  {\rm for} \  \  j=l+1/2  \  \  .
\eea

\ni
For the 1$S_{1/2}$ electrons, the quantum numbers are
$n=1$, $l=0$ and $\kappa=-1$, and a good approximation for the
ground state eigenfunctions is
\bea
&&F_{1,-1}=N\sqrt{m \lp 1+s   \rp} \ r^{s-1} e^{-\gamma m r} \nn \\
&&G_{1,-1}=-N\sqrt{m \lp 1-s   \rp} \ r^{s-1} e^{-\gamma m r}  \  ,
\eea

\ni
where
\beq
\gamma=Ze^2
\eeq

\ni
and
\beq
s=\sqrt{1-\gamma^2}  \  .
\eeq

\ni
The normalization is made using 
\bea
\int \psi^* \psi \ d^3x &=& \int_0^\infty \lp |F|^2+|G|^2  \rp r^2dr= \nn \\
&=&  2N^2m^2 
\int_0^\infty r^{2s}e^{-2\gamma mr} dr = 1 \ ,
\eea

\ni
that determines the constant
\beq
N={1\over m} {\lp 2\gamma m \rp^{s+1/2} \over \sqrt{2\Gamma(2s+1)}}  \  .
\eeq

\ni
So,
\bea
\langle r\rangle &=& 2N^2m^2 
\int_0^\infty r^{2s+1}e^{-2\gamma mr} dr = 
{2s+1\over 2 \gamma m }
      \nn \\
\langle r^2\rangle&=& 2N^2m^2
\int_0^\infty r^{2s+2}e^{-2\gamma mr} dr = 
{(2s+2)(2s+1)  \over (2 \gamma m)^2 } \nn \  \    ,   \\
\eea

\ni
and from these results,
\beq
\Delta r={\sqrt{2s+1}\over2\gamma m}
\eeq

\ni
that gives
\beq
r_{min}(Z)= {2.6459226.10^4\over Z}
\lp 2\sqrt{1-\alpha^2Z^2} +1 
-\sqrt{2\sqrt{1-\alpha^2Z^2}}   \rp 
\   {\rm fm}  \  \   .
\label{rmin}
 \eeq

\ni
Another value of interest, is the value of $r$ where the wave function
reaches the maximum, that may be calculated using
\beq
{d(rF)\over dr}=0={\sqrt{1-\alpha^2Z^2}\over r}- \alpha m Z   \  \  , 
\eeq

\ni
that gives
\beq
r_{peak}= {\sqrt{1-\alpha^2Z^2}\over \alpha mZ}    \  ,
\eeq

\ni
and from $G$, one gets the same relation.

\ni
The results for $r_0$, $r_{min}$ and $r_{peak}$ are shown in Fig. 1-3.

In this framework, 
atomic instability effects are expected to occur in elements that present
 $r_{min}<r_0$, what means that the
1$S_{1/2}$ electron has significant probability of reaching the
horizon of events. In this case, the ground state is unstable, and from
(\ref{rmin}), one notes that it 
 happens for  $Z>92.39$.
In view of these results (see Fig. 1)
there is no surprise in the fact that elements with
atomic numbers greater than the critical value 
$Z$=92 (Uranium) are not found in Nature, and exists
only if synthesized artificially by man. If $r_0=r_{peak}$, the element will be
very unstable, and it happens for
\beq
Z\sim {0.618\over\alpha}\sim 107.73
\eeq

\ni
that is a little bit above the last element, Rf (Rutherfordium), that has 
$Z$=104, fact that shows one more time that the theory
is in agreement with the observable world.

A point that must be remarked,
is that these results are obtained only studying the electronic
structure of the elements,
 without considering their nuclear structures. As it is well known,
 some aspects of the
interaction responsible for the shell structure of the nuclei need to be
understood, and one factor that may be considered, specially for heavy
elements,
 is the effect of instability
presented in this work.

\begin{figure}[hbtp] 
\centerline{
\epsfxsize=9.cm
\epsffile{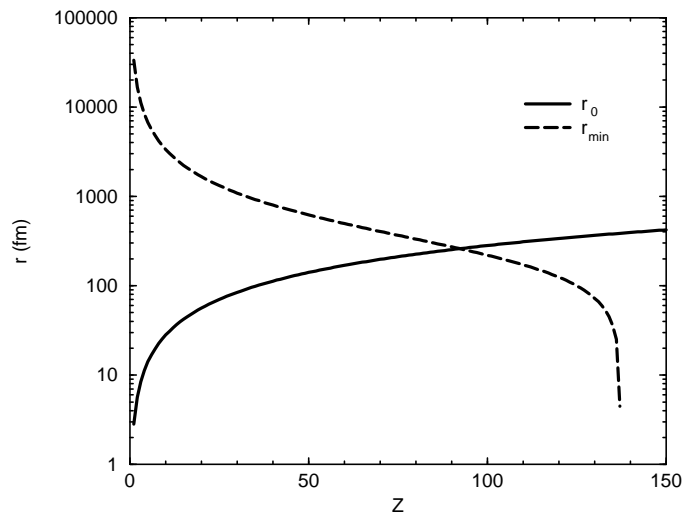}}
\caption{Comparison between $r_0$ and $r_{min}$ as functions of the nuclear 
charge, $Z$.}
\end{figure}

\begin{figure}[hbtp] 
\centerline{
\epsfxsize=8.cm
\epsffile{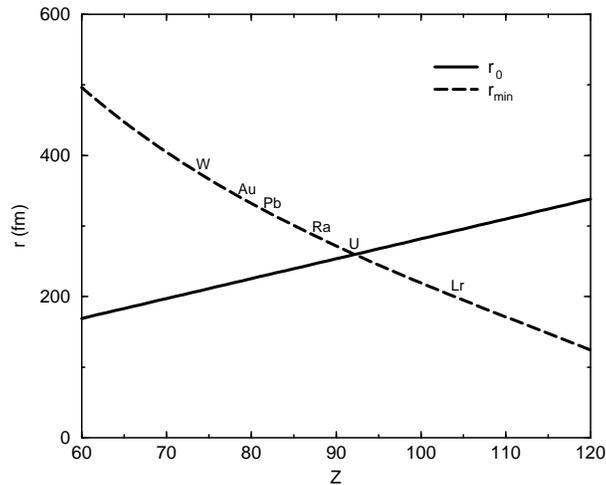}}
\caption{Comparison between $r_0$ and $r_{min}$. }
\end{figure}

\begin{figure}[hbtp] 
\centerline{
\epsfxsize=8.cm
\epsffile{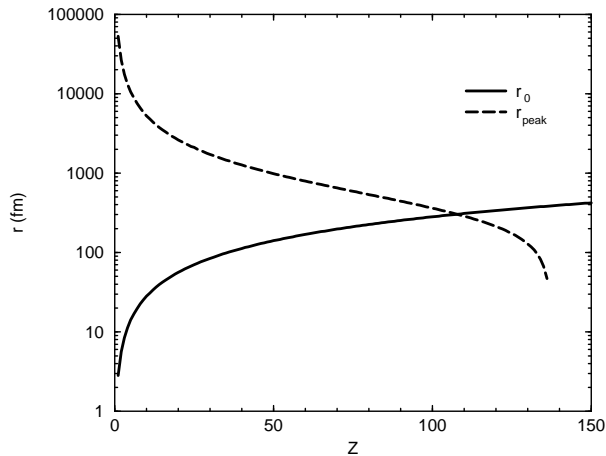}}
\caption{Comparison between $r_0$ and $r_{peak}$ as functions of the nuclear 
charge, $Z$.}
\end{figure}

\section{Summary and conclusions}

In this paper the study of the general quantum mechanics 
is been continued.
The metric is determined by the interaction of quantum objects, such as
electrons and quarks. The effect of the
energy-momentum tensor of electromagnetic and strong
interactions in the space-time has been considered, and 
with this procedure the 
the constant of the field equations has been calculated.
The results obtained are consistent with the ones of \cite{cb1}.

One can observe the presence of the mass in the constant $A$, what  does
not happen in the general relativity, fact that is due to the equivalence 
principle. Observing that
 $A\propto g/m^2$, one concludes that the effect of the curvature
of space-time, for a particle of mass $m$ in a field,
 decreases for large masses and increases for small masses
and for large coupling constants. It is interesting to note that 
a dynamical cut-off is determined by this theory, as it was used in  eq. 
(\ref{mmm}), providing correct results.

Another observation that must be made is that
initially, spherical symmetry has been supposed, but eq. (\ref{bfld})
 is general, independent of the symmetry of the system.
This equation may also be used with the inclusion of other interactions,
as the Yang-Mills one, for example, considering the 
Yang-Mills field tensor $F_{\m\n}^a$ in the construction of the energy-momentum
tensor, and quark confinement, from the results of this theory,
is expected to  occur.

One must remark that with the development of the general quantum
mechanics, we are being able 
to explain many characteristics of the studied physical systems \cite{cb1},
\cite{cb2}, 
using the new proprieties that appears from the understanding
of the geometry of the space-time.
The instability effect presented in this paper is another result of the theory
that does not appear in the usual quantum mechanics, despite the fact
that it is an observable effect.

The Dirac theory introduced the special relativity in quantum mechanics,
so it is very reasonable to think that the next step is to formulate the 
quantum mechanics
in a way analogous of the one that the general relativity is. But in fact,
the step proposed here, lead to a theory that absolutely 
  is not a quantized version of general relativity,
many differences occur and some concepts of the general relativity are not the
same ones in this formulation. 
The atomic spectrum obtained in this way shows  that
the corrections of the energy levels, due to this  general formulation of
quantum mechanics  with the inclusion of the 
electric interaction in  the space-time metric,
 provide a quite impressive 
agreement with the experimental data,
and is a strong evidence in 
 the  validation the theory. 
Another interesting result is the hadron model \cite{cb2}, where quark
confinement is shown to be a consequence of this theory.

\begin{acknowledgments}
I wish to thank M. R. Robilotta and Y. Hama for useful discussiosss.

\end{acknowledgments}


\end{document}